\documentclass[12pt]{article}
\usepackage{graphicx}
\usepackage{epsfig}
\begin{document}

\begin{center}
{\bfseries {\large The QCD analysis of $xF_3$ structure function based on the analytic approach}}
\vskip 5mm
{A.V. Sidorov$^{\dag}$ and O.P. Solovtsova$^{\dag,\ddag}$}
\vskip 5mm
{\small {\it $^\dag$ Joint Institute for Nuclear Research, 141980 Dubna, Russia}} \\
{\small {\it $^\ddag$ Gomel State Technical University, 246746 Gomel, Belarus}}
\end{center}

\begin{abstract}
We apply analytic perturbation theory to the QCD analysis of the
$xF_3$ structure function data of the CCFR collaboration. We use
different approaches for the leading order $Q^2$ evolution of the
$xF_3$ structure function and compare the extracted values of
the parameter $\Lambda$ and the shape of the higher twist
contribution. Our consideration is based on the Jacobi polynomial
expansion method of the unpolarized structure function.
The analysis shows that the analytic approach provides
reasonable results in the leading order QCD analysis.
\end{abstract}

\section{Introduction}

The data on the $xF_3$ structure function \cite{CCFR97} provide a
possibility for a precise test of the perturbative QCD
predictions for the $Q^2$ evolution of this structure function.
The analysis of $xF_3$ data is simplified because one does not need to
parameterize gluon and see quark contributions and can
parameterize the shape of the  $xF_3$ structure function  itself
at some value $Q^2_0$. For the kinematics region of these data
$Q^2 $ $\geq 1.3$ GeV$^2$ the higher twist (HT) contribution to
the structure function should be taken into account.
This allows us to study from the above-mentioned data both the
perturbative part and the HT correction related to each other.
Here, we will focus our attention on the interplay of the different
approaches to the strong coupling $Q^2$-behavior and the
$x$-dependence of the HT contribution.

In our investigation we apply the analytic approach in QCD
proposed by Shirkov and Solovtsov \cite{DVS}, the so-called
analytic perturbation theory (APT) (see also
Refs.~\cite{KS1,SolSh:99-TMP}). This method takes into account
the basic principles of local quantum field theory which in the
simplest cases is reflected in the form of $Q^2$-analyticity of
the K\"all$\acute{{\rm e}}$n--Lehmann type.
The key point of APT constructions---the analytic properties of
some functions (the two-point correlator of the quark currents, the
moments of the structure functions and so on). An overview of the
analytic approach to QCD can be found in Ref.~\cite{ShSol06}.
In the framework of the APT, in contrast to the infrared behavior
of the perturbative (PT) running coupling, the analytic coupling
has no unphysical singularities. At low $Q^2$ scales, instead of
rapidly changing $Q^2$ evolution as occurs in the PT case, the APT
approach leads to slowly changing functions (see, e.g.,
Refs.~\cite{GLS-our,KPSST12}). In the asymptotic region of large
$Q^2$ the APT and the PT approaches coincide.
It should be noted that the moments of the structure functions
should be analytic functions in the complex $Q^2$ plane with a cut
along the negative real axis (see Ref.~\cite{JLD-00} for more
details), the ordinary PT description violates analytic properties
due to the unphysical singularities of PT coupling. On the other
hand, the APT supports these analytic properties.
For fullness,  in our analysis, we consider also the recent
variant of the model for the freezing-like behaviour coupling --
``massive analytic perturbative QCD'' (MPT)~\cite{MPT} (see
Ref.~\cite{Yndurain,KotKri10} for a discussion).

In Refs.~\cite{BMS05,BMS06} further development of the APT
method was made -- the generalization to the fractional powers of the
running coupling which is called the Fractional Analytic Perturbation
Theory (FAPT) (see Ref. \cite{Bakulev} as review). The FAPT
technique was applied to analyze the $F_2$ structure function
behavior at small $x$-values \cite{KotKri10,CIKK09}, to analyze the
low energy data on nucleon spin sum rules $\Gamma_1^{p,n}(Q^2)$
\cite{PSTSK09}, to calculate binding energies and masses of
quarkonia~\cite{AC13}.
Here, we continue applications of the APT/FAPT approach
executing the data on the $xF_3$ structure function and investigating
how the analytic approach works in this case by comparison with the
standard PT analysis.

\section {The Method of the QCD analysis}

In our analysis, we will follow the well-known approach based on the
Jacobi polynomial expansion of structure functions. This method of
solution of the DGLAP equation was proposed in Ref.~\cite{PS} and
developed for both unpolarized  \cite{Kretal} and polarized cases
\cite{LSS}. The main formula of this method allows an approximate
reconstruction of the structure function through a finite number
of Mellin moments of the structure function
\begin{equation}
xF_{3}^{N_{max}}(x,Q^2)=\frac{h(x)}{Q^2}+x^{\alpha}(1-x)^{\beta}
\sum_{n=0}^{N_{max}}
\Theta_n ^{\alpha , \beta}
(x)\sum_{j=0}^{n}c_{j}^{(n)}{(\alpha ,\beta )}
M_{j+2}       \left ( Q^{2}\right ).
\label{Jacobi}
\end{equation}
The $Q^2$-evolution of the moments $ M_N(Q^2)$ in the leading
order (LO) perturbative QCD is defined by
\begin{equation} \label{m3q2}
M^{QCD}_N(Q^2)
= \left [ \frac{\alpha _{s}\left ( Q^{2}\right )}
{\alpha _{s}\left ( Q_{0}^{2}\right )}\right ]^{{\frac{\gamma^{(0),N}}{2\beta_0}}}
M_N^{QCD}(Q^2_0) ,~~~N = 2,~3, ... \,.
\end{equation}
Here ~$\alpha_s(Q^2)$~ is the QCD running coupling,
~$\gamma^{(0),N}$~ are the  nonsinglet leading order anomalous dimensions,
$\beta_0=11-2n_f/3$ is the first coefficient of the renormalization
group $\beta$-function, and $n_f$ denotes the number of active flavors.

Unknown coefficients $M_N^{QCD}(Q^2_0)$ in Eq.~(\ref{m3q2}) could be parameterized
as the Mellin moments of some function:
\begin{equation}
M_3^{QCD}(N,Q^2_0)=\int_{0}^{1}dx{x^{N-2}}Ax^{a}(1-x)^{b}(1+\gamma~x),
~~ N = 2,3, ... ~~.
\label{Mellf30}
\end{equation}

The shape of the function $h(x)$ as well as parameters A, $a$, $b$, $\gamma$,
and $\Lambda_{\rm{QCD}}$  are found by fitting
the experimental data to the $xF_3(x,Q^2)$ structure function
\cite{CCFR97}.
The detailed description of the fitting procedure could be found in
Ref.~\cite{KKPS2}. The term $h(x)/Q^2$ is considered as pure
phenomenological. The target mass corrections are taken into
account to the order $o(M^4_{nucl}/Q^4)$.

\section{Analytic approach in QCD}

The APT method gives the possibility of combining the
renormalization group resummation with correct analytic properties
in $Q^2$-variable for some physical quantities and provides
also a well-defined algorithm for calculating higher-loop
corrections~\cite{SolSh:99-TMP}.
As the difference between the APT and PT running couplings becomes
significant at low $Q^2$-scales (see, e.g., Fig.~1 in
Ref.~\cite{GLS-our}) this stimulates applications of the analytic
approach to a new analysis \cite{ShSol06},
especially after the generalization of the APT to the fractional
powers of the running coupling (see Refs.~\cite{Bakulev,CK-2012,Stefanis}
for further details).

In the framework of the analytic approach the following
modification in the standard PT expression (\ref{m3q2}) for the
$Q^2$-evolution of the moments $ M_N(Q^2)$ is required: $\left[
\alpha_{\rm PT}(Q^2) \right]^{\nu} \Rightarrow$
${\cal{A}}_{\nu}(Q^2)$. It transforms Eq.~(\ref{m3q2}) as
follows\footnote{Beyond LO see Refs.~\cite{KotKri10,A-MS}.}:
\begin{equation}
{\cal{M}}^{QCD}_N(Q^2)
 = \frac{{\cal{A}}_{\nu}(Q^{2})}
{{\cal{A}}_{\nu}(Q_{0}^{2})} \,
{\cal{M}}_N^{QCD}(Q^2_0) ~, ~~ {\nu} \equiv
{\frac{\gamma^{(0),N}}{2\beta_0}}
\, , \label{m3q2APT}
\end{equation}
where the analytic function ${\cal{A}}_{\nu}$ is derived from the
spectral representation and corresponds to the discontinuity of the
$\nu$-th power of the PT running coupling
\begin{equation}
\label{del_APT}
{\cal{A}}_{\nu}(Q^2) \,=\,\frac{1}{\pi}\,
\int_0^\infty\,\frac{d\sigma}{\sigma\,+\,Q^2} \,
 {\rm Im} \;
\left \{ {\alpha}_{\rm PT}^{\nu}(-\sigma -{\rm i}\varepsilon)
 \right\}\,  .
\end{equation}
Note that the function ${\cal{A}}_{1}(Q^2)$ defines the APT running
coupling: ${ \alpha}_{\rm APT}(Q^2)$ $ \equiv
{\cal{A}}_{1}(Q^2)\,$. The mathematical tool for numerical calculations
of ${\cal{A}}_{\nu}$ for any $ \nu >0 \,$ up to four-loop order in
the perturbative running coupling is given in Ref.~\cite{KB13}.

The `normalized' analytic function $\bar{\cal{A}}_{\nu}= \beta_0
{\cal{A}}_{\nu} /(4 \pi)$ in the leading order (LO) has rather
a simple form (see, e.g., \cite{Bakulev}) and can be writhen as
\begin{eqnarray}
 && \bar{\cal{A}}_{\nu}^{LO}(Q^2)=\left[\bar{a}^{LO}_{\rm PT}(Q^2)\right]^{\nu}\,- \,
\frac{{\rm {Li}_\delta}(t)}{\Gamma(\nu)} \,, \label{a_nu}  \\[0.2cm]
 && ~ {\rm {Li}}_{\delta}(t)= \sum_{k=1}^{\infty}
 \frac{t^k}{k^{\delta}} , ~~ t=\frac{\Lambda^2}{Q^2},~~~\delta=1-\nu,~\nonumber
\end{eqnarray}
where the `normalized' PT running coupling $\bar{a}^{LO}_{\rm
PT}(Q^2)=\beta_0 {{\alpha}_{\rm
PT}^{LO}(Q^2)}/(4\pi)=1/\left[\ln(Q^2/\Lambda^2)\right]\,$ and
Li$_\delta$ is the polylogarithm function. For ${\nu=1}$
expression (\ref{a_nu}) leads to the well-known one-loop APT
result~\cite{DVS}
\begin{equation}
\label{a_1}
 \alpha_{\rm APT}^{LO}(Q^2)={\alpha}_{\rm {PT}}^{LO}(Q^2) +
 \frac{4 \pi}{\beta_0}\, \frac{\Lambda^2}{\Lambda^2-Q^2} \, .
\end{equation}

One could see that at large $Q^2$ the second term in 
the r.h.s. of expression (\ref{a_1}) is negative. It was confirmed qualitatively in the
phenomenological analysis of the $xF_3$ data in
Ref.~\cite{sid1Q2}.

\begin{figure}
\begin{center}
{\includegraphics[width=7.5cm]{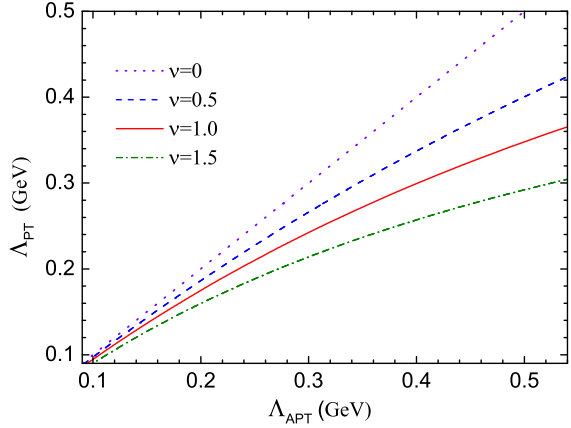}} \caption {The behavior
of the parameter $\Lambda_{{\rm PT}}$ vs. $\Lambda_{{\rm APT}}$  in
LO for different values of $\, \nu \,$ at $Q_0^2=3$~GeV$^2$.}
\label{Fig_1-Sidorov}
     \end{center}
\end{figure}

It should be stressed that values of the QCD scale parameter
$\Lambda$ are different in the PT and APT approaches. In order to
illustrate this, in Fig.~1, we present the behavior following from
the condition  $ \left[ \alpha_{\rm PT}^{LO}(Q_0^2,\Lambda_{{\rm
PT}}) \right]^{\nu} = {\cal{A}}_{\nu}^{LO}(Q_0^2, \Lambda_{{\rm
APT}})$ of the parameter $\Lambda_{{\rm PT}}$ vs. $\Lambda_{{\rm
APT}}$ for different values of $\, \nu \,$.

In short, one-loop modification of the QCD coupling within the MPT
approach which will be considered further corresponds to the
replacement of the logarithm in the ${\alpha}_{\rm
{PT}}^{LO}(Q^2)$ to the ``long logarithm'' with the ``effective
gluonic mass''\footnote{The parameter of the
``effective mass" serves as an infrared regulator and typically of
the order $m_{gl} =500 \pm 200$~MeV (see, e.g.,
 Ref.~\cite{m_gl}).}  $m_{gl}$: $\ln(Q^2/\Lambda^2) \Rightarrow \ln [
(Q^2 + m_{gl}^2) /\Lambda^2]$ (see,
Refs.~\cite{MPT,Cvetic-MPT}).

\section {Numerical analysis of experimental data}

The results of the LO QCD fit in different approaches are
presented in Table 1 and Figs. 2--5. Both cases $h(x)$--~free and
$h(x)=0$ are considered for $Q^2_0=3$~GeV$^2$, $Q^2 >
1.3$~GeV$^2$, $n_f=4$, and $N_{Max}=12$. In order to reconstruct
the $x$-shape of the HT contribution, we have parameterized $h(x)$
in the number of points $x_i$ = 0.015, 0.045, 0.080, 0.125, 0.175,
0.225, 0.275, 0.35, 0.45,  0.55, 0.65 - one per $x$-bin. The
values of A, $a$, $b$, $\gamma$, $x_i$ and $\Lambda$ are considered
as free parameters.
\begin{table}[!t] \small\small
 \caption{The results for the QCD  leading order
fit (with TMC) of $xF_3$ data \cite{CCFR97}
($Q^2_0=3$~GeV$^2$, $Q^2 > 1$~GeV$^2$, $n_f=4$, and $N_{Max}=12$). }
\begin{center}\label{tab:1}
 {\begin{tabular}{|l|c|c|c|c|c|}   \hline
 \multicolumn{1}{|c}{~~~~}  &
 \multicolumn{2}{|c|}{$h(x)=0$} &
 \multicolumn{2}{c|}{$h(x)$--free } \\ [1mm] \hline
Approach &  $\Lambda$ (MeV) &
~~$\chi^2_{d.f.}$ & $\Lambda$ (MeV) & ~~$\chi^2_{d.f.}$ \\ [1mm]  \hline  \hline
~~ PT   & $291\pm 36$ & $1.35$ & $363\pm 170$ & $0.984$  \\
~~ APT   & $275\pm 39$ & $1.42 $& $350\pm 145$ & $0.980$  \\
~~ MPT  & $299\pm 38$ & $1.35$ & $351\pm 128$ & $0.985$  \\
``naive" analyt.   & $417\pm 83$ & $1.34 $& $412\pm 240$ & $0.980$  \\
 \hline
\end{tabular} } \end{center}
\end{table}

\begin{figure}[t]
\begin{minipage}{.46\textwidth}
\centerline{\includegraphics[width=1.03\textwidth]{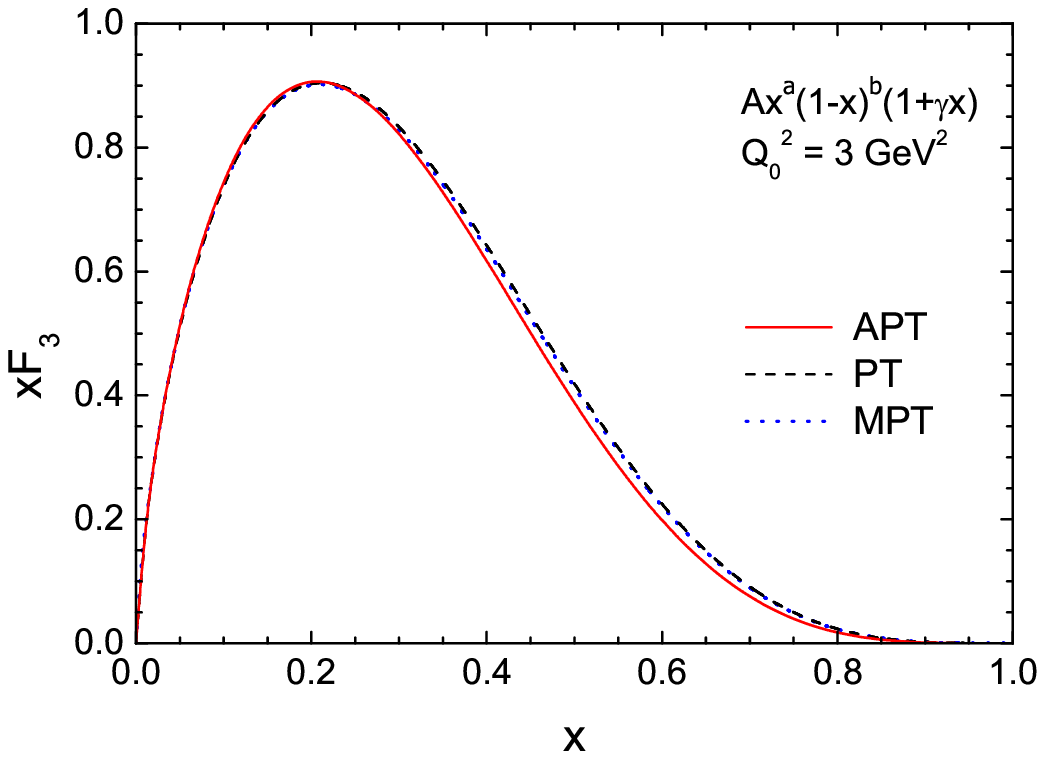}}
\caption {Comparison of  parametrizations of $xF_3$ in the PT,
APT and MPT approaches for $h(x)=0$.} \label{Fig_2-SS}
\end{minipage}
\phantom{}\hspace{0.5cm}%
\begin{minipage}{.46\textwidth}
\centering
\includegraphics[width=1.03\textwidth]{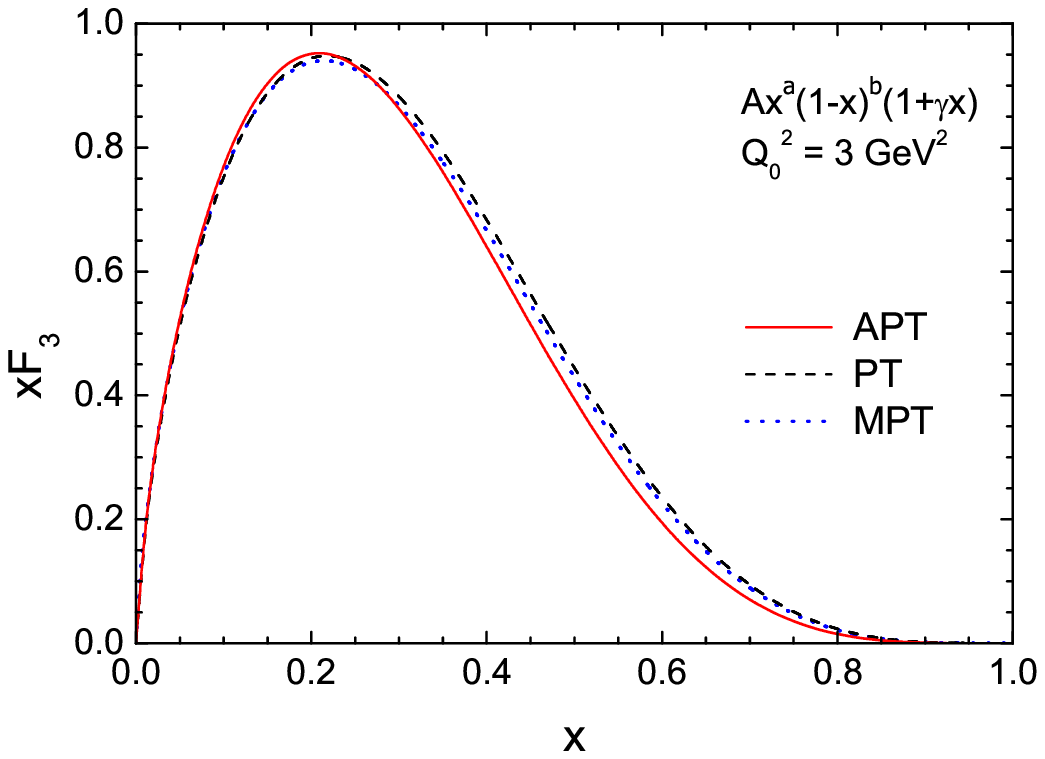}
\caption {Comparison of parametrizations of $xF_3$ in the PT,
APT and MPT approaches for $h(x)\neq0$. } \label{Fig_3-SS}
\end{minipage}
\end{figure}

As can be seen from Table~1, the values of parameter the $\Lambda$ for the
case  $h(x)=0$ are smaller in comparison with the case of nonzero
HT contribution. The difference of the $\Lambda$ values for
APT and PT are smaller in the analysis with the HT contribution:
$(\Lambda_{{\rm PT}}-\Lambda_{{\rm APT}})_{h(x)=0}>(\Lambda_{{\rm
PT}}-\Lambda_{{\rm APT}})_{h(x)\neq0}$. The LO $h(x)=0$ results
for $\Lambda$ values are consistent within errors.
If one adds the HT contribution, the values of parameter the $\Lambda$
and their errors are higher than  $h(x)=0$ case.

For illustrative purposes we present in the last line of
Table~1 the result corresponding to the use in the analysis of
``naive analytization" when the ordinary perturbative coupling is replaced
by the analytic coupling: $ {\alpha}_{\rm {PT}}(Q^2) \to
\alpha_{\rm APT}(Q^2)$
(see Ref.~\cite{BMS05} and references
therein).

\begin{figure}
\begin{minipage}{.47\textwidth}
\centerline{\includegraphics[width=1.0\textwidth]{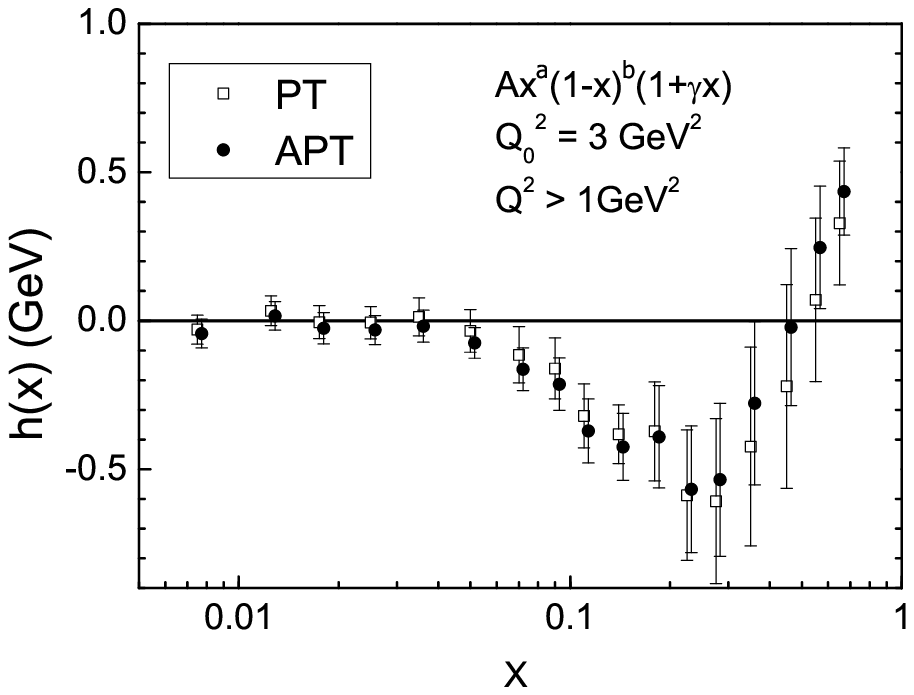}}
\caption {Higher twist contribution resulting from the LO QCD analysis  of
$xF_3$ data \cite{CCFR97} for the PT and APT approaches. }
\label{Fig_4-Sidorov}
\end{minipage}
\phantom{}\hspace{0.5cm}%
\begin{minipage}{.47\textwidth}
\centering
\includegraphics[width=1.01\textwidth]{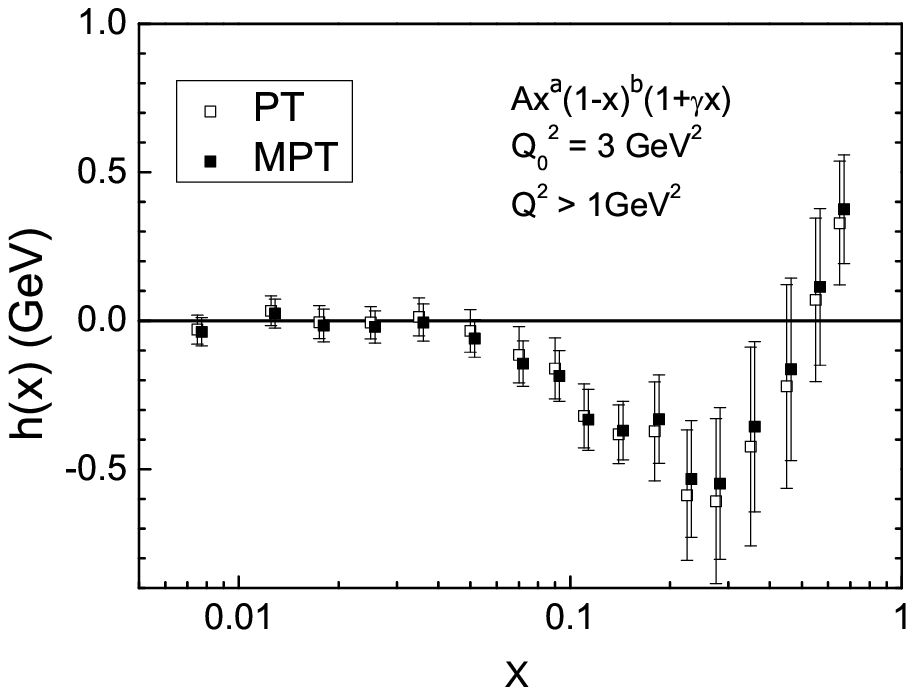}
\caption {Higher twist contribution resulting from the LO QCD analysis  of
$xF_3$ data \cite{CCFR97} for the PT and MPT approaches. }
\label{Fig_5-Sidorov}
\end{minipage}
\end{figure}

Figures 2--3 show the $xF_3$-shape obtained in the APT, PT and
MPT approaches  without taking into account the HT term (Fig.~2) and with
the HT (Fig.~3).
In both cases, the result for the APT approach is slightly higher
than for the PT and MPT ones for small $x$ and less for large $x$.

Figures 4--5 demonstrate
the shape of the HT contribution. From Fig.~4 one can see
that  for $x>0.3$ we obtained $h^{APT}(x)
> h^{PT}(x)$. This inequality is in qualitative agreement with the
result obtained in LO for the shape of the HT contribution for
the non-singlet part of the $F_2$ structure function (see Table~3 in
Ref.~\cite{KotKri10}). The opposite inequality is obtained for small values
$x<0.2\,$: $h^{APT}(x) < h^{PT}(x)$.  Figure~5
shows that the central values of $h^{PT}(x)$ and $h^{MPT}(x)$ are
very close to each other.

\section{Conclusion}
We performed the QCD analysis of the $xF_3$ structure function data
based on the analytic approach. It should be noted that the wide
kinematic region experimental points gave us the possibility to
analyze HT contributions of both small and relatively large $x$
and to compare the APT and MPT results to the PT one. We have found
that in the examined region $Q^2 > 1$~GeV$^2$ the values of
$\Lambda$ obtained in the PT, APT and MPT approaches are close to each
other, while the ``naive  analytization" approach leads to a
rather large $\Lambda$ value. The shape of the HT contributions is
in quantitative agreement with the results of the previous
analysis of the  $xF_3$ structure function data.
We made the first step --  LO analysis which
showed that the analytic approach gives reasonable results. It is
important to extend the analysis to higher orders and apply it
to the structure function data in the low $Q^2$ region.

\vspace{0.3cm}
{\bf Acknowledgments}
\vspace{0.1cm}

It is a pleasure for the authors to thank V.L. Khandramai, S.V.~Mikhailov,
and O.V.~Teryaev for interest in this work and helpful discussions.
This work was partly supported by the RFBR grants 11-01-00182 and 13-02-01005 and the
BelRFFR-–JINR grant F12D-002.


\end{document}